\documentclass[fleqn,twoside]{article}
\usepackage{espcrc2,psfig}

\usepackage{graphicx}
\usepackage[figuresright]{rotating}

\def\ltsima{$\; \buildrel < \over \sim \;$}
\def\lsim{\lower.5ex\hbox{\ltsima}}
\def\gtsima{$\; \buildrel > \over \sim \;$}
\def\gsim{\lower.5ex\hbox{\gtsima}}

\newcommand{\be}{\begin{equation}}
\newcommand{\en}{\end{equation}}
\newcommand{\ergs}{\rm \ erg \; s^{-1}}

\def\cmdue {\rm \ cm^{-2}}

\def\msole {~M_{\odot}}

\newcommand{\AmS}{{\protect\the\textfont2
  A\kern-.1667em\lower.5ex\hbox{M}\kern-.125emS}}

\title{New results on neutron star low mass transients in the quiescence}

\author{S. Campana\address[OAB]{INAF -- Osservatorio astronomico di Brera, 
        Via Bianchi 46, \\ 
        I--23807 Merate (LC), Italy}%
        \thanks{e-mail: campana@merate.mi.astro.it}
        L. Stella\address[OAR]{INAF -- Osservatorio Astronomico di Roma,
        Via Frascati 33,\\ I--00040 Monteporzio Catone (Roma), Italy}
}
       
\begin{document}

\begin{abstract}
We review the main observational properties of neutron star 
low mass transients in quiescence. We first survey the discoveries of BeppoSAX.
We then focus on recent discoveries by Chandra and XMM-Newton, with special
emphsasis on the detection of the quiescent counterpart of SAX J1808.4--3658
and the study of variability in the quiescent state.  
\vspace{1pc}
\end{abstract}

\maketitle

\section{Introduction}

Many Low Mass X--ray Binaries (LMXRB) accrete matter at very
high rates, and therefore shine as bright X--ray sources only sporadically.
Among transient sources are Soft X--ray Transients (SXRTs) host an
old neutron star orbiting a low mass star companion (for a review see Campana
et al. 1998a). 
These systems usually alternate periods (weeks to months) of high X--ray
luminosity, during which they share the same properties of persistent LMXRBs, 
to long (1--10 years) intervals of quiescence in which the X--ray luminosity
drops by up to 5--6 orders of magnitude.

\begin{figure*}[htb]
\centerline{\psfig{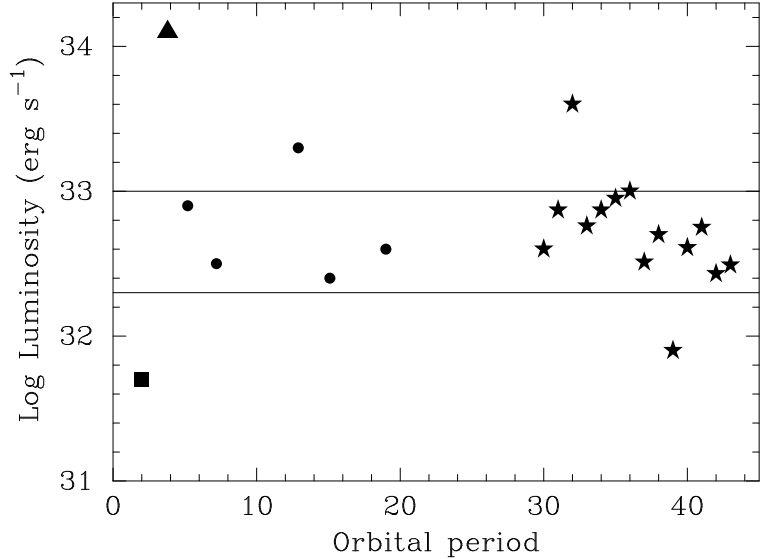}}
\caption{Quiescent 0.5--10 keV (unabsorbed) luminosity of a sample of SXRTs. 
The triangle is EXO 0748--673 and it is peculiar since it is not clear if it
is a true transient system. The five filled circles are well known systems
(e.g. Aql X-1, Cen X-4, etc.). The filled square is SAX J1808.4--3658. The
stars are globular clusters SXRTs in quiescence recently discovered by
Chandra. Their orbital periods are false.}
\label{sxrt_lum}
\end{figure*}

In recent years the quiescent properties of a handful SXRTs have been studied 
in some detail thanks to dedicated campaigns carried out with BeppoSAX and
RXTE during outburst and observations in quiescence with XMM-Newton and Chandra. 
Among the main outcomes of these investigations are: 1) SXRT quiescent
luminosities are for most cases within a narrow range of
$10^{32}-10^{33}\ergs$ (0.5--10 keV, See Fig. \ref{sxrt_lum}); 2) an increasing 
number of candidate SXRTs in quiescence is being discovered in globular
clusters observed with Chandra; 3) the quiescent X--ray spectra of SXRTs
display a soft component and in some cases a hard (power-law) component
contributing $10-50\%$ of the flux in the 0.5--10 keV band.

Observationally, the soft component can be modelled with a black body of
0.1--0.3 keV temperature and few km radius. Especially promising is the idea
that the soft component of SXRTs may be produced from the cooling of the
neutron star heated during repeated outbursts (van Paradijs et al. 1987;
Stella et al. 1994; Campana et al 1998a). The theory of deep crustal heating
by pycnonuclear reactions compares well with the observations (Brown et
al. 1998; Campana et al. 1998a; Rutledge et al. 1999; Colpi et al. 2001). In
particular, Rutledge et al. (1999) fitted neutron star atmospheric models to
the soft component of quiescent spectra of SXRTs and derived temperatures in
the 0.1--0.3 keV and radii consistent with the neutron star radius (10--15
km). 

The hard component is well described by a power law tail.
In the quiescent spectrum of Aql X-1 and Cen X-4 observed by ASCA and BeppoSAX
this component is statistically significant (Asai et al. 1996, 1998; Campana
et al. 1998b, 2000) with photon index in the 1--2 range. The same power law
is needed (even if less significant statistically) in the analysis of Chandra
data in order to achieve an emitting radius of the cooling component
consistent with the neutron star radius (otherwise the inferred radius would
be smaller; Rutledge et al. 2001a, 2001b). In some sources in globular
clusters instead there is no indication off the presence of a hard tail
(e.g. Heinke et al. 2003), as well as in a observation of 4U 1608--52 in
quiescence with ASCA (Asai et al. 1996). 
The nature of this hard component is still uncertain. Models range from
Comptonization in a hot corona to Advection/Convection Dominate Accretion Flow
(ADAF/CDAF; Menou et al. 1999) to emission from a low luminosity jet (Fender
et al. 2003) or shock emission from the neutron star that resumed its radio
pulsar activity in quiescence. 
The latter model envisages a situation similar to that of the eclipsing
radio pulsar PSR B1259--63 or of the `black widow' pulsar PSR B1957+20:
a shock at the boundary between the relativistic MHD wind from the radio
pulsar and the matter outflowing from the companion star 
(Tavani \& Arons 1997; Tavani \& Brookshaw 1991;
Campana et al. 1998a). For the model to explain the observed luminosity in the
hard power law component of quiescent SXRTs, a few percent
of the pulsar spin-down luminosity must be converted into shock
emission. Ongoing deep searches in the radio band have not yet
revealed any steady or pulsed emission from quiescent SXRTs (Burgay et
al. 2003); however free-free absorption due to matter in the
binary system and its surroundings might be an important limiting factor in
these searches (Stella et al. 1994; Burgay et al. 2003). 
%\cite{Scho70,Mazu84}.

\section{History}

\begin{figure*}[htb]
\leftline{\psfig{figure=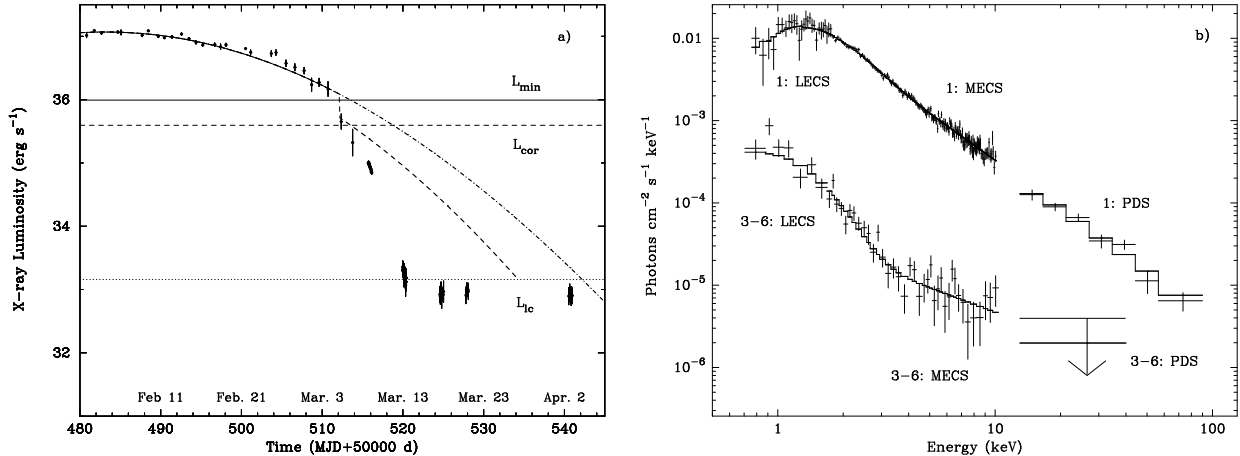,width=4.5truecm}}
\caption{Light curve of the Feb.-Mar. 1997 outburst of Aql X-1 (panel {\it a}). 
Data before and after MJD 50514 were collected with the RXTE ASM 
(2--10 keV) and the BeppoSAX MECS (1.5--10 keV), respectively.
The evolution of the flux from MJD 50480 to MJD 50512 is well fit 
by a Gaussian centered on MJD 50483.2. This fit however does not provide 
an acceptable description for later times (see the dot-dashed line), not even
if the accretion luminosity is calculated in the propeller regime 
(dashed line). 
The straight solid line represents the X--ray luminosity corresponding to
the closure of the centrifugal barrier $L_{\rm min}$
(for a magnetic field of $10^8$ G and a spin period of 1.8 ms)
and the straight dashed line the luminosity gap due to the action of the
centrifugal barrier, $L_{\rm cor}$. The dotted line marks the minimum 
luminosity in the propeller regime ($L_{\rm lc}$).
Panel {\it b} shows the BeppoSAX unfolded spectra of Aql X-1 during the early 
stages of the fast decline (1) and during the quiescent phase (3--6, summed). 
The best fit spectral model (black body plus power law) is
superposed to the data.}
\label{aql}
\end{figure*}

March-April 1997 BeppoSAX observations of Aql X-1 were the first to
monitor the evolution of the spectral and time variability
properties of a neutron star SXRT from the outburst decay to quiescence.
A fast X--ray flux decay was observed, which brought the source luminosity
from $\sim 10^{36}$ to $\sim 10^{33}\ergs$ in less than 10 days (see Fig. \ref{aql}). 
This behaviour was later been observed also in other SXRTs as well as Aql X-1
itself.  

The X--ray spectrum showed a pronounced hardening in correspondence of the
X--ray flux turn-off (Zhang et al. 1998; Campana et
al. 1998b). This result has been confirmed and extended with RXTE data of the
2002 outburst (Campana et al. 2003).  
At a luminosity of $\sim 10^{35}\ergs$ the X--ray spectrum showed a power 
law high energy tail with photon index $\Gamma\sim 2$ up to 100 keV, markedly
different from the thermal spectrum  extending up to 30--40 keV observed at
luminosities a factor of 10 higher.
At even lower luminosities the X--ray spectrum became even harder with
$\Gamma \sim 1-1.5$. 
These observations, together with the detection by RXTE of a periodicity
of a few milliseconds during an X--ray burst, likely indicate that the rapid
flux decay is caused by the onset of the propeller effect arising from the very
fast rotation of the neutron star magnetosphere (Campana et al. 1998b; Zhang
et al. 1998) and that a further transition may occur approaching quiescence
(Campana et al. 1998b). This transition may underline the re-activation of a
millisecond radio pulsar in quiescence state of Aql X-1, the relativistic wind
of which powers the quiescent X--ray emission.

\section{SAX J1808.4--3658}

\begin{figure*}[thb]
\vskip -8truecm
\centerline{\psfig{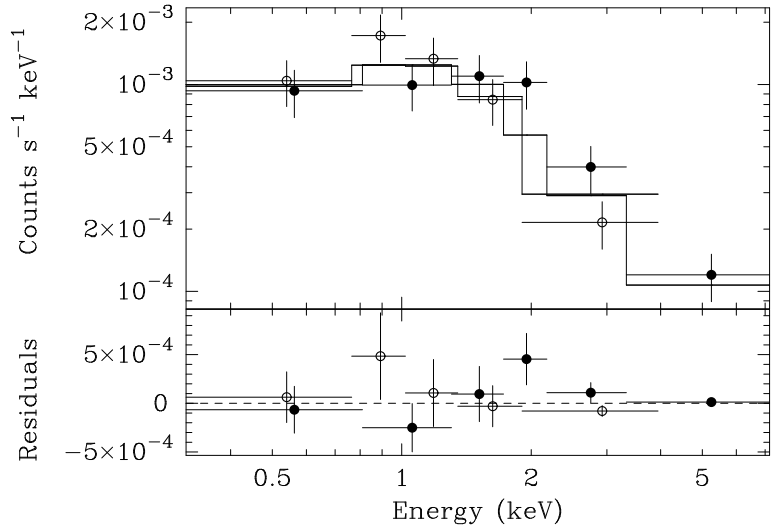}}
\caption{MOS1 (open dots) and MOS2 (filled dots) spectrum of SAX J1808.4--3658.
Overlaid is the fit with an absorbed power law model described
in the text. In the lower panel are reported the residuals of the fit.}
\label{1808}
\end{figure*}

The study of LMXRBs has been revolutionized by the discovery of the first
source (the SXRT SAX J1808.4--3658) showing coherent millisecond pulsations in
the persistent emission of its outburst state (Wijnands \& van der Klis
1998). Other four millisecond pulsating transients have later been discovered. 

SAX J1808.4--3658 showed the same decrease to quiescence as Aql X-1 (Gilfanov
et al. 1998) with a hard power law appearing at luminosities of $\sim
5\times10^{35}\ergs$ (Gilfanov et al. 1998). 
A few attempts to reveal SAX J1808.4--3658 in quiescence have been carried
out with ASCA and BeppoSAX (Stella et al. 2000; Wijnands et al. 2002a).
These observations revealed a source at a luminosity of $\sim 10^{32}\ergs$;
yet the field is crowded, source confusion a problem for the $\sim 1$ arcmin
resolution achieved by BeppoSAX and ASCA and no firm conclusions could be
drawn. 
An XMM-Newton observation with a high signal to noise revealed SAX J1808.4--3658
in quiescence isolated, but a nearby source (about 1 arcmin) can have altered
the position and the flux observed with previous instruments (Campana et
al. 2002).  
The 0.5--10 keV unabsorbed luminosity is
$5\times10^{31}\ergs$, a relatively low value compared with other neutron
star SXRTs. Moreover, at variance with other SXRTs, the quiescent spectrum 
of SAX J1808.4--3658 was dominated by a hard ($\Gamma\sim 1.5$) power law 
with only a minor contribution ($\lsim 10\%$), if any, 
from a soft black body/neutron star atmosphere component (Fig. \ref{1808}). 
If the power law originates in the shock between the wind of a turned-on radio
pulsar and matter outflowing from the companion, then a spin-down to X--ray
luminosity  conversion efficiency of $\eta\sim 10^{-3}$ is derived; this is in
line  with the value estimated from the eclipsing radio pulsar PSR
J1740--5340 (D'Amico et al. 2001). 
Within the deep crustal heating model, the faintness of the blackbody-like 
component indicates that SAX J1808.4--3658 may host a massive neutron
star ($M\gsim1.7\msole$). In fact this is the only way to have a rapid
cooling thanks to direct Urca process and in turn neutrino cooling.

\begin{figure*}[htb]
\centerline{\psfig{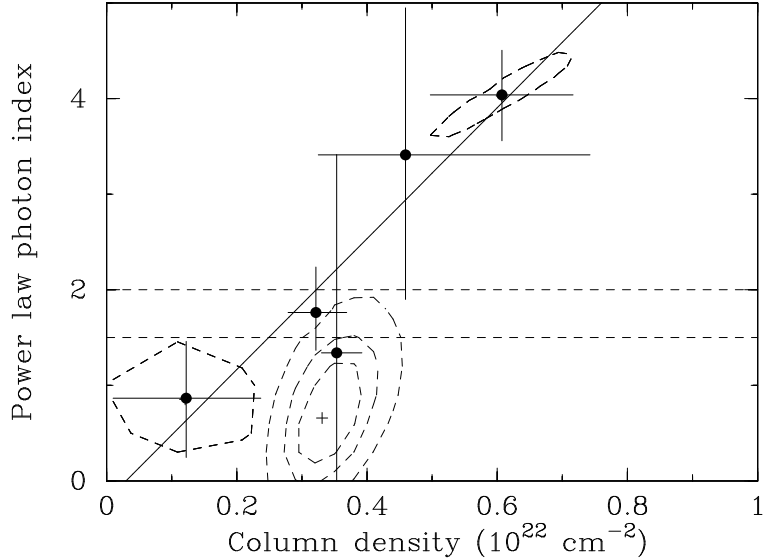}}
\caption{Power law photon index vs. column density correlation of the five Aql
X-1 observations. Overplotted is the best linear fit. Dashed lines indicate
the range over which the synchrotron emission model likely applies. On the
hardest and softest observations $1\,\sigma$ countours have been superposed. 
The 1, 2, $3\,\sigma$ countours obtained fitting the entire set of data with a 
single power law and column density is also reported.}
\label{aql_corr}
\end{figure*}

\begin{figure*}[htb]
\centerline{\psfig{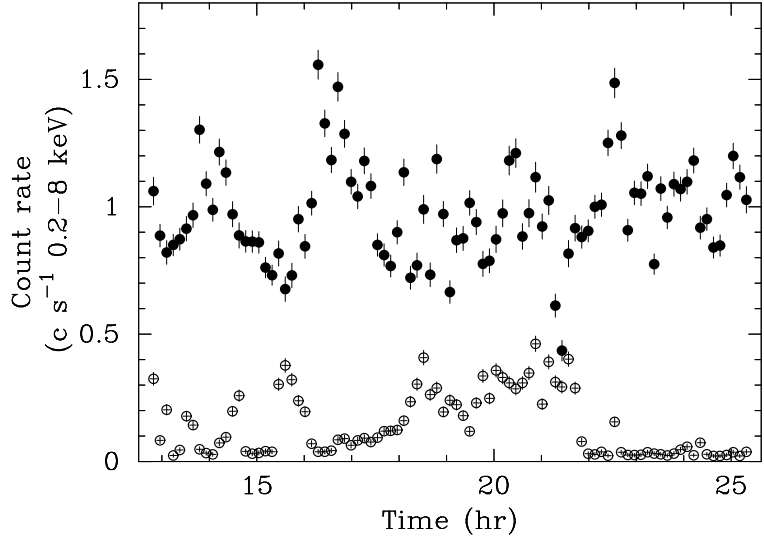}}
\caption{Background subtracted 0.2--8 keV pn light curve of Cen X-4. The bin
size is 500 s. Open dots represent the background light curve scaled to the
source extraction area (which has been already subtracted in the light curve above).}
\label{cen_short}
\end{figure*}
 
\begin{figure*}[htb]
\centerline{\psfig{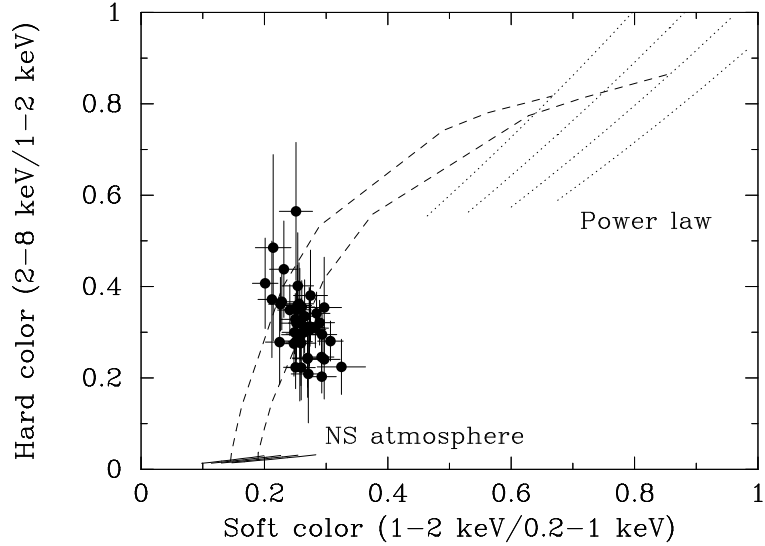}}
\caption{Color-color diagram of the pn background subtracted light curve. The
soft color is defined as the ratio of the counts in the 0.2--1 keV to 1--2 keV,
the hard color as the ratio of the counts in the 2--8 keV to 1--2 keV. On top
of this we depict hardness ratios for an absorbed power law model (with photon 
index from 1.2 to 2) and an absorbed neutron star atmosphere model (with
temperatures from 70 to 100 eV) for different values of column densities (3,
5, 7, $9\times 10^{20}\cmdue$, dotted lines). Two different (dashed) curves
connect the single component models with increasing contribution from the two
for column densities of 5 and $9\times 10^{20}\cmdue$.}
\label{cen_color}
\end{figure*}
 
\section{Variability in quiescence}

Short term variability is potentially a powerful tool for the study of the emission
mechanism(s) responsible for the SXRTs quiescent emission. 
A factor of 3 variability over timescales of days (Campana et al. 1997) 
and $40\%$ over 4.5 yr (Rutledge et al. 2001b) was reported in Cen X-4.
Several other neutron star systems have also been found to be
variable in quiescence by factors of 3--5 (e.g. Rutledge et al. 2000) but data
have been collected over several years and with different instruments.
Interesting results have been recently found thanks to the observations of well
known systems with Chandra and XMM-Newton.

After years of persistent emission at a high level, KS 1731--260 turned to
quiescence (Wijnands et al. 2001; Burderi et al. 2002). The source was
observed first with Chandra and 6 months later with XMM-Newton. The quiescent
luminosity decrease by a factor of 2--5 (Wijnands et al. 2002b). In the cooling neutron star
model, this decrease implies that the crust of the neutron star cooled down
rapidly between the two epochs, indicating that the crust has a high
conductivity. Further monitoring of KS 1731--260 in quiescence can provide
crutial information on the crust conductivity and level of impurities
(Rutledge et al. 2002a). A similar turn-off of the X--ray luminosity after
years of activity occurred in X1732--304 (in Terzan 1; Wijnands et al. 2002c)
and MXB 1659--29 (after 2.5 yr of activity, Wijnands et al. 2002d). These
sources may represent a new population of long-lasting transients.

Rutledge et al. (2002b) analyzing Chandra data of the Aql X-1 quiescent phase 
after the November 2000 outburst found a variable flux and X--ray spectrum.
They interpreted these variations in terms of variations of the neutron star 
effective temperature, which changed from $130^{+3}_{-5}$ eV, down to
$113^{+3}_{-4}$ eV, and finally increased to $118^{+9}_{-4}$ eV.
Interestingly, during the last observation they also found short-term ($<10^4$
s) variability (at $32\%$ rms).
These data are the first to show an increase of the quiescent flux in the
quiescent SXRT. The latter result has a direct impact on the quiescent 
emission models, since no known mechanism associated with crustal heating can
account for this variability. Rutledge et al. (2002b) suggested that accretion might
occur during the Aql X-1 quiescent state and variations have can be
ascribed to a variable mass inflow rate.

Campana \& Stella (2003) reanalyzed the Chandra spectra of Aql X-1, together
with a long BeppoSAX observation in the same period, and propose a different 
interpretation of the spectral variability: that this is due to correlated
variations of the power law component and the column density ($>5$, a part of which
might be intrinsic to the source), while the temperature and flux of the
neutron star atmospheric component remained unchanged (Fig. \ref{aql_corr}). 
The power law slope vs. column density behaviour is in qualitative agreement
with that observed from the radio pulsar binary PSR 1259--63, lending support
to the idea that the power law component arises from emission at the shock
between a radio pulsar wind and inflowing matter from the companion star.

Thanks to XMM-Newton large throughput, Cen X-4 was observed with the
highest signal to noise ever. A $\sim 13$ hr observation revealed rapid ($>100$ s), 
large ($45\pm7\%$ rms in the $10^{-4}-1$ Hz range) intensity variability,
especially at low energies (Fig. \ref{cen_short}). In order to
investigate the cause of this variability, Campana et al. (2003) divided the
data into intensity intervals and fit the resulting spectra with the 
canonical model for neutron star transients in quiescence, i.e. an absorbed
power law plus a neutron star atmosphere model. 
Variations across different spectra can be mainly accounted for by a variation
in the column density together with another spectral parameter
(Fig. \ref{cen_color}). Based on the available 
spectra, a variation of the power law could not be preferred over a variation in
the temperature of the atmosphere component (even if the first is slightly
better in terms of reduced $\chi^2$). This variability can be accounted for by
accretion onto the neutron star surface (e.g. Rutledge et al. 2002b) or by the
variable interaction between the pulsar relativistic wind and matter
outflowing from the companion in a shock front (e.g. Campana \& Stella 2003). 

\begin{table*}[!htb]
\begin{center}
\caption{\label{highbase} Standard spectral models fits for the high and low
count rate interval.}
\begin{tabular}{cccccc}
\hline\hline
Model    & $N_H$               &$k\,T$   &Photon index         & $\chi^2_{\rm red}$& N.h.p.  \\
         &($10^{20}$ cm$^{-2}$)& (eV)    &(hard component)     &     (d.o.f.)         &         \\
\hline
%fix
Low      & $6.0\pm2.9$       &$87\pm1$ &$1.53\pm0.17\ (40\%)$& 2.60 (450)        &\ $0\%$ \\
Medium   &                   &         &                     &                   &        \\
High     &                   &         &                     &                   &        \\
%10.8\pm0.4
\hline
%nhfree
Low      & $15.1\pm3.6$      &$78\pm9$ &$1.60\pm0.14\ (35\%)$& 1.51 (448)        &\ $0\%$ \\
Medium   & $8.9\pm4.3$       &         &                     &                   &        \\
High     & $4.4\pm3.4$       &         &                     &                   &        \\
%15.8\pm5.3
\hline
%nhfree+atmo
Low      & $5.1\pm2.6$       &$86\pm7$ &$1.47\pm0.13 \ (46\%)$& 1.02 (446)        & $37\%$ \\
Medium   & $4.6\pm2.6$       &$91\pm5$ &$ \hskip 2.2cm (38\%)$&                   &        \\
High     & $6.0\pm2.1$       &$99\pm7$ &$ \hskip 2.2cm (30\%)$&                   &        \\
%9.7\pm2.4
\hline
%nhfree+pow
Low      & $9.6\pm4.6$      &$85\pm6$  &$1.32\pm0.19 \ (33\%)$& 1.03 (444)        & $34\%$ \\
Medium   & $5.4\pm3.7$      &          &$1.62\pm0.16 \ (42\%)$&                   &        \\
High     & $4.6\pm3.8$      &          &$2.11\pm0.20 \ (53\%)$&                   &        \\
%11.5\pm6.0
%========
\hline\hline
\end{tabular}
\end{center}

\noindent In the case of a varying neutron star atmosphere model the radius
$9.7\pm2.4$ km  (intrinsic radii), whereas for a varying
power law we have a radius of $11.5\pm6.0$ km.

\end{table*}

\section{Conclusions}

The study of SXRTs in quiescence is revealing a rich phenomenology. For the
interpretation of this, different emission mechanism(s) and underlying
different physical scenarios are actively debated. These depend on the
influence of the neutron star rotation and magnetic field on the infalling
matter as well as the physics of deep crustal heating during intense accretion
episodes. 
%Thus, SXRTs
%represent a unique testbed for plasma physics at low accretion rates.
In the last few years, mainly after the discovery of coherent pulsations in
the transient SAX J1808.4--3658, a large interest in these sources has grown.
Chandra and XMM-Newton (quiescence) as well as BeppoSAX and RXTE (outburst)
observations yielded many interesting results. In particular, the discovery of
a sizeable population of candidate SXRTs in globular clusters is providing important new
clues on the statistical properties of SXRTs in quiescence.
Besides the most studied sources, peculiar behaviours are being found in other
sources such as SAX J1808.4--3658 itself, characterized by a factor of 2 lower mean
luminosity and, more importantly, a spectrum dominated by the power law component.

On the other hand, in depth studies of as well known sources with Chandra and
XMM-Newton are still providing new and unexpected results. In the case of Aql
X-1 monitoring its quiescent state with Chandra led to the discovery of
luminosity and spectral variability over a month timescale. This can either
been interpreted as accretion 
occurring at these low rates (with an uninfluent magnetic field, Rutledge et
al. 2002b) or as correlated variability in the quantity of matter around the
system and change in the power law slope due to the interaction of a
relativistic pulsar wind with this matter (Campana \& Stella 2003).
In the case of Cen X-4 an XMM-Newton observation disclosed rapid ($\lsim 100$
s) X--ray variability, the nature of which, while still uncertain, can be
accounted for by the two different mechanisms discussed above (Campana et al. 2003).

\end{document}